\documentstyle[epsf]{mn}

\def\Rg{R_{\rm g}}
\def\dm{{\dot m}}

\def\Rin{R_{\rm in}}

\def\mdot{\dot m}
\def\MSun{M_{\odot}}
\def\Zycki{$\dot{\rm Z}$ycki}
\def\Rozanska{R\'o$\dot{\rm z}$a\'nska }
\def\Lubinski{Lubi\'nski }
\def\Gierlinski{Gierli\'nski }

\title[J1550--564]
{Mapping the inner accretion disk of the galactic black hole J1550--564 
through its rise to outburst}
\author[Colin D. Wilson and C. Done]
    {Colin D. Wilson and Chris Done \\
        University of Durham, Department of Physics, 
       South Road, Durham DH1 3LE; c.d.wilson@durham.ac.uk, chris.done@durham.ac.uk \\
}

\begin{document}

\maketitle

\begin{abstract}

We study the spectral properties of the first 14 observations of the
rise to outburst phase of the X--ray transient J1550--564. Using both
the PCA and HEXTE instruments, we find that the 3--200 keV spectra
smoothly pass from a standard low/hard state to a very high state. The
classic high state is never encountered possibly indicating that it is
not a phenomenon of the rise phase. We find that the individual PCA
spectra can be fitted adequately by a disk black body and a thermal
Comptonization model which includes reflection. Once the very high
state is reached there is clear spectral curvature of the continuum which
possibly indicates the presence of a composite thermal/non-thermal
plasma.

Our detailed modelling of the reflection parameters shows a sharp
increase in mean ionization at the onset of the transition between
the low state and very high state. There is a related variability in
the reflected fraction but its exact value depends on the continuum
model used. The reflected fraction varies around values of
$\Omega/2\pi \sim 0.1$ and is never
consistent with $\Omega/2\pi = 1$. We can
constrain the inner radius using relativistic smearing and while there
are large uncertainties, the data are incompatible with a disk
extending to the last stable orbit ($6R_G$) in either state. 

Since the system is on the rise to outburst, the disk instability models
(and observed increasing QPO frequency) strongly imply that there is no
standard inner disk at the time the low state spectrum is observed. This
is compatible with a truncated disk, filled by an X--ray hot, advection
dominated accretion flow.  However, magnetic flares above the outbursting
disk can also match the observed spectra once the effects of either
outflow and/or strong photoionization of the surface of the disk are
included. We clearly see strong ionization of the reflector in the very
high state. This is probably from collisional ionization, as the disk
surface temperature is $\sim 0.7$ keV. This can strongly suppress
reflection from the inner disk. 

\end{abstract}


\section{Introduction}

Accreting galactic black hole candidates (GBHCs) offer the most direct
method of investigating accretion physics. They are free of magnetic
fields and surface boundary layers and so have the simplest possible
accretion flow. Many of these systems are also transient, showing huge X--ray
outbursts where the luminosity rises by many orders of magnitude over a
short time and then fades away more slowly. The evolution of their spectra
and variability during these outbursts can be used to determine the nature
of the accretion flow as a function of mass accretion rate onto the
central object. 

It is now generally believed that the outbursts are caused by a classic
disk instability. In quiescence the material in the disk is cool, so the
sound speed is slow, and the mass accretion rate through the disk is
smaller than the mass transfer rate from the companion star. The material
then builds up on the outer edge of the disk and the structure is very
different from that of a steady state disk (Shakura \& Sunyaev 1973,
hereafter SS).  When the local temperature in any part of this quiescent
disk reaches $\sim 10^{4}K$, hydrogen begins to ionize. This results in a
sudden increase in opacity, causing an increase in temperature and the
local sound speed, and so giving an increase in local mass accretion rate.
The disk interior to this point then receives this higher mass accretion
rate, which can trigger a heating wave that propagates inwards (and
outwards). This switches the disk into the hot state, which is well
described by an SS disk but with a mass accretion rate which is much
higher than that from the companion star. Over time, this depletes the
disk, draining it sufficiently to switch back into the cool state (see e.g
the review by Osaki 1996, with applications to the black hole transients
in King \& Ritter 1998). 

This picture fails to explain the observed spectra during both
outburst and quiescence. The SS disk models assume that the gravitational
potential energy of the accreting matter is dissipated in optically thick
material, so it thermalizes with a maximum temperature of $\sim 1$ keV. In
quiescence the disk is very cool, $\sim 4000$ K independent of radius
(e.g. Cannizzo 1998), and so emits no high energy flux. Yet observations
show that these systems generically produce hard X--ray emission. At high
mass accretion rates (approaching Eddington) the spectra are dominated by
a soft component at $kT\sim 1$ keV with a strongly (very high state: VHS)
or weakly (high state: HS) Comptonized component (which is occasionally
below the detection limit). The Comptonized component forms a rather steep
power law tail ($\Gamma\sim 2-3$) which extends out beyond $511$ keV in
the few objects with good high energy data (Grove et al. 1998).  At lower
mass accretion rates, below $\sim 5$ per cent of Eddington, there is a
rather abrupt transition when the soft component drops in temperature and
luminosity. Instead this (low state: LS) spectrum is dominated by thermal
Comptonization, with $\Gamma < 1.9$, rolling over at energies of $\sim
150$ keV (see e.g. the reviews by Tanaka \& Lewin 1995; van der Klis 1995;
Nowak 1995).  At even lower luminosities, during quiescence, there is weak
hard X--ray emission which appears to be similar to the LS spectrum (e.g.
Kong et al. 2000). 

There are several possible ways to produce the hard X--rays, but all
require that some fraction of the accretion energy is dissipated in an
optically thin environment. The two main candidates currently considered
are magnetic flares above the disk, generated by the Balbus--Hawley MHD
dynamo responsible for the disk viscosity, or that the inner disk is
replaced by an optically thin, X--ray hot accretion flow. Little is known
about the spectrum expected from electrons heated through magnetic
reconnection, but the observed weak disk emission in the LS can be
explained if most of the viscous dissipation is released in a patchy or
out-flowing magnetic corona (Svensson \& Zdziarski 1994; Stern et al. 1995;
Beloborodov 1999). By contrast, the properties of an optically thin
accretion flow can be worked out in some detail, typically giving electron
temperatures of $\sim 100$ keV (cooling dominated by radiation: Shapiro,
Lightman \& Eardley 1976; or advection: Narayan \& Yi 1996). These can
exist only at fairly low mass accretion rates as they rely crucially on
the assumption that the accretion energy is given mainly to the protons,
and that the electrons are only heated via Coulomb collisions. At high
densities (i.e. high mass accretion rates) then the electrons efficiently
drain energy from the protons and the flow collapses back into an SS disk.
Hence these flows can contribute only to the hard X--rays in LS or
quiescent emission. The collapse of such flows may give the physical
mechanism for the LS/VHS state transition (Esin, McClintock \& Narayan
1997). In the HS and VHS then magnetic reconnection is an obvious
candidate for the X--ray tail. However, there is also a radio jet/outflow
seen in the LS and VHS which may also play a role in the X--ray emission
(Fender 2000). 

In summary, for quiescence and LS there are two potential mechanisms for
the hard X--ray production. One of these has a truncated disk, and an
inner optically thin, hot flow, while the other has a (steady state or
quiescent) disk which extends down to the last stable orbit. If there is a
truncated disk in quiescence/LS, then the disk material has to {\it move}
in order to give the HS and VHS spectra which are dominated by emission
from an optically thick, inner disk. Data from the rise to outburst of a
GBHC transient would then give one of the best diagnostics of the nature
of the accretion flow as the majority of the disk material is {\it known}
to be moving inwards in this phase.

Such data have only recently become possible with the rapid response of
the RXTE satellite. RXTE J1550--564 is the first GBHC to be intensively
observed during its rise to outburst phase, as well as its decline, giving
unprecedented coverage of the X--ray spectrum over a huge range in mass
accretion rate. These data give several ways to observationally track the
disk. Firstly there are the Quasi--Periodic Oscillations (QPO's) seen in
the X--ray spectra.  While the origin of these are not well understood,
{\it all} QPO models use a characteristic radius (this radius is a 
local change in disk properties and is sometimes associated with the
inner edge of an SS disk, e.g Psaltis \&
Norman 2001). For the QPO frequency to change then requires that this
radius is {\it not} fixed at the last stable orbit of $6\Rg$ (where
$\Rg=GM/c^2$). RXTE J1550--564 shows a dramatic increase in QPO frequency
during the rise, by a factor of $\sim 50$ (Cui et al. 1999, Remillard
et al. 1999), most probably
indicating that the transition radius (or SS disk radius) is moving
inwards during the rise (di Matteo \& Psaltis
1999)

A second way to track the inner disk is to use X--ray reflection. Hard
X--rays illuminating optically thick material give rise to a Compton
reflection component and associated iron fluorescence line (Lightman \&
White 1988; George \& Fabian 1991; Matt, Perola \& Piro 1991). These
features are smeared by special and general relativistic effects of the
motion of the disk in the deep gravitational potential well (Fabian et al.
1989). The amount of reflection gives the solid angle of the optically
thick disk as observed from the hard X--ray source, while the amount of
smearing shows how far the material extends into the gravitational
potential of the black hole. 

Here we analyze the detailed spectral properties during the rise phase. 
The 3-200 keV spectral changes during this time are shown in Figure
\ref{fig:ls_vhs}. The 2--20 keV spectrum showed a dramatic softening
during the rise (See the light curve and hardness ratio plots in
Sobczak et al. 1999), concurrent with a change in the variability power
spectrum (Cui et al. 1999). Taken together, these imply a LS/VHS state
transition during the rise. We examine the PCA and HEXTE spectra, showing
that the broadband continuum is indeed consistent with a LS/VHS
transition, and identify a smeared reflected component in all the spectra.
For the LS, the reflection signature is similar to that seen in other LS
spectra, i.e. reflection from mainly neutral material, which subtends a
solid angle of substantially less than $2\pi$ and is broadened, but not by
as much as expected if the reflecting material extended down to the last
stable orbit (\Gierlinski et al. 1997; \Zycki, Done \& Smith 1997; Done \&
\Zycki \ 1999; Gilfanov et al. 1999; Zdziarski, \Lubinski \& Smith 1999).
Given that both the disk instability model and QPO data imply a truncated
SS disk at this point then this strongly favours models which identify the
LS with a truncated disk, and rules out models which {\it require} an
inner disk (of the Shakura and Sunyaev form) for the source of the hard
X--ray emission in the LS. 

\section{The rise to outburst data from RXTE J1550--564}

Calibration issues are important in any detailed spectral analysis. We
have attempted to quantify these by analyzing a Crab spectrum taken
contemporaneously with the RXTE J1550--564 rise data.  The Crab spectrum
is complex, since it contains both the nebula and pulsar components. The
synchrotron nebula gives a spectrum which is approximately a power law
($\Gamma=2.3$) in the 3--20 keV band, although there is subtle
spectral curvature over a wider bandpass (e.g. Atoyan \& Aharonian
1996). The pulsar spectrum is
generically harder but has stronger curvature and is phase variable. It is
consequently only approximately a power law over a rather restricted
energy band (see e.g. Pravdo, Angelini \& Harding 1997; Massaro et al.
2000 and references therein). We approximate the total spectrum in the
3--20 keV range by two power laws, representing the nebula and pulsar,
respectively.  We fix the pulsar component to a spectral index of
$\Gamma=1.8$ and its flux at 10--20 \% of that of the nebula, and obtain a
good fit ($\chi^2_\nu=1.2-1.0$) using only the top layers from detectors 0
and 1, with $0.5 \%$ systematic errors. 

Given the complex nature of the total emission from the Crab, the HEXTE
data should not lie on a single power law extrapolation of the PCA. This
is an important point, since this `discrepancy' has been used to argue
that the PCA--HEXTE cross calibration is unreliable (Sobczak et al. 1999).
While work on the cross calibration is still ongoing, there is currently
no reason not to fit the two datasets together, allowing for a
normalization offset. 

\begin{figure}
\epsfxsize=\hsize
\begin{center}
\epsfbox{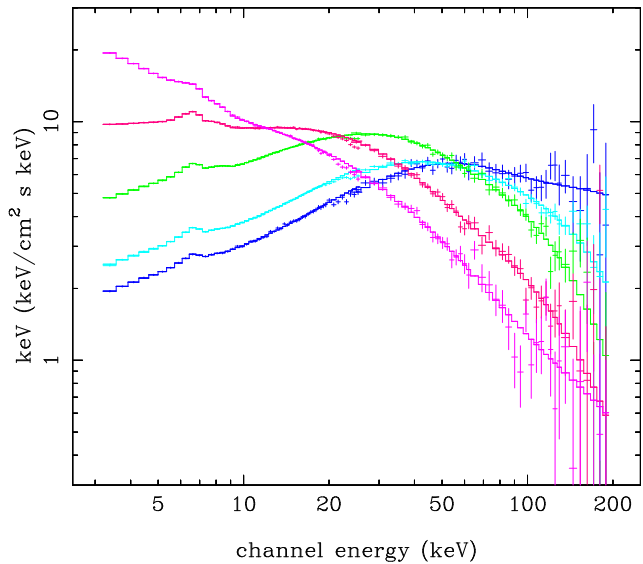}
\end{center}
\caption{The spectral transition of XTE J1550--564 from low/hard to
very high state.}
\label{fig:ls_vhs}
\end{figure}

We have used the public RXTE dataset 30188-06, which traces the rise of
the outburst from 07/09/98 to the 16/09/98. The PCA (detectors 0,1 top
layer only) and HEXTE (detector 0 only) data were extracted using the REX
script in FTOOLS 5.0. Systematic errors of 0.5\% were added to all the PCA
spectra. Selected spectra from the rise are shown in Figure
\ref{fig:ls_vhs}, showing the dramatic spectral softening. Although the
2-10 keV light curve of the ASM shows a spectacular increase in luminosity
(a factor of $\sim 7$), figure \ref{fig:ls_vhs} shows that this is a
bandpass effect caused by ignoring the higher energies. The estimated
bolometric luminosity only increases by a factor of $\sim 2$. 

\section{model description}

We analyzed the spectra using the XSPEC spectral fitting package
(Arnaud 1996) v10.0. The spectrum is expected to contain soft emission
from the accretion disk, which we model using the {\tt diskbb} and
hard emission from Compton scattering of these soft seed photons by
energetic electrons.  A power law with exponential rollover at the
electron energy is often used to approximate this, but it seriously
overestimates the Comptonized flux at energies close to those of the
seed photons, and underestimates it below the cutoff. This renders
most parameters derived from such models unreliable. Instead we use an
approximate solution for thermal Comptonization based on Kompaneets'
equation ({\tt thComp}: Zdziarski, Johnson \& Magdziarz
1996). Inherent limitations on the Kompaneets' equation mean that this
solution becomes inaccurate at optical depths $<1$ and temperatures
above $>100$ keV.  The Comptonized spectrum can illuminate the
accretion disk, giving rise to a reflected (reprocessed)
component. Both the continuum reflection (Magdziarz \& Zdziarski 1995)
and iron emission line are calculated self--consistently for a given
ionization state (\Zycki \ \& Czerny 1994), and the total reprocessed
spectrum is then relativistically smeared by convolving it with the
{\tt diskline} model of Fabian et al.  (1989) including corrections
for light bending. The reflected spectrum is then determined by the
solid angle of the reflector as seen from the X--ray source, $\Omega$,
normalized to $2\pi$, its inclination $i$, ionization parameter,
$\xi=4\pi F_{X}/n$ (where $F_{X}$ is the illuminating flux and $n =$
number density),
and inner disk radius, $\Rin$ (see \Zycki \ ,Done and Smith 1999 for model
details). The abundances of Morrison \& McCammon (1983) are used.

There are as yet no good determinations of inclination of the system, so
we fix this at $30^\circ$. The distance is also fairly uncertain, but
E(B--V)$\sim 0.7$ measured from optical spectra suggests 2.5 kpc
(Sanchez--Fernandez et al. 1999). This E(B--V) predicts a column density of
$\sim 4\times 10^{21}$ cm$^{-2}$, consistent with the column we
measure from fitting the publicly
available archived ASCA GIS data (sequence ID 15606000 taken on 12/09/98).
We fix the column to this value in all the fits, and note that the much
larger column of $\sim 2\times 10^{22}$ cm$^{-2}$ inferred by Sobczak et
al. (1999) is an artifact of their more approximate spectral model. 

\section{results}
\subsection{Initial spectrum}

The PCA data alone cannot be well fit by a thermal Comptonization
continuum and disk blackbody spectrum ($\chi^2_\nu=304.3/40$). Including
a reflected spectrum and its associated iron fluorescence line
gives a good fit to the data ($\chi^2_\nu=43.0/38$), but this is further
{\it significantly} improved by including relativistic smearing
($\chi^2_\nu=37.1/37$, for $R_{in}=26^{+63}_{-15}$). The reflected
fraction is $\Omega/2\pi=0.24^{+0.03}_{-0.04}$ from mainly neutral material,
$\xi=16^{+23}_{-5}$. The disk blackbody is {\it not} significantly
detected in the spectrum, and assuming that it provides the soft
photons for the thermal Comptonization gives a limit on its
temperature of $\le 0.5$ keV). 

These findings are qualitatively different to those derived from using
more approximate spectral models. The often used continuum of a disk
blackbody and power law, together with a broad Gaussian line and smeared
edge to phenomenologically model the reflected spectral features (as used
for later parts of the outburst in Sobczak et al. 1999) gives a comparably
good fit if the column is allowed to be free
($\chi^2_\nu=26.0/36$). With this model the disk blackbody is {\it
required}, and has
$kT=0.75^{+0.12}_{-0.11}$ keV. We strongly caution against using
parameters derived from such phenomenological fits to derive physical
quantities such as disk radii (see also the detailed criticism of the {\tt
diskbb} model of Merloni, Fabian \& Ross 2000). 

\begin{figure}
\epsfxsize=\hsize
\begin{center}
\epsfbox{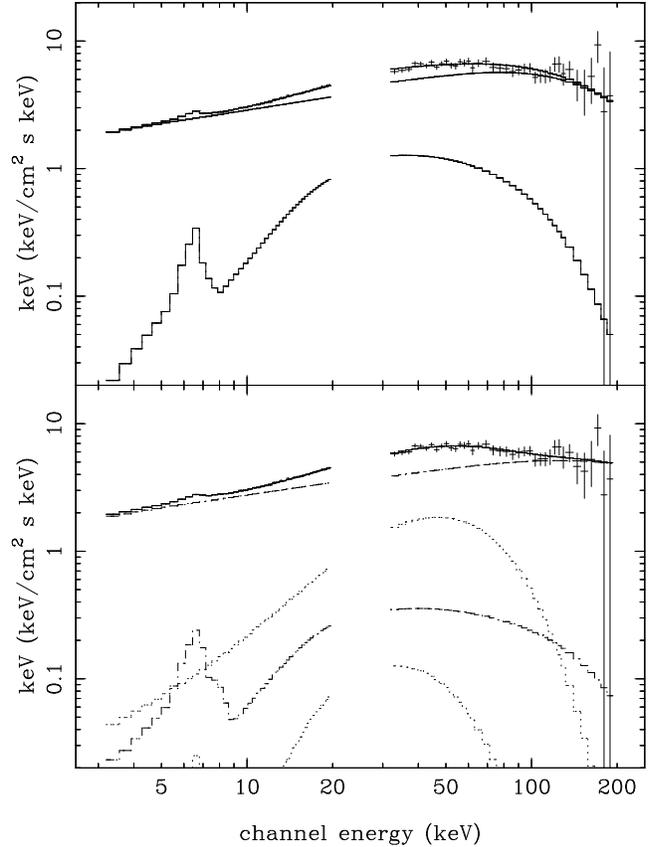}
\end{center}
\caption{The spectra of J1550--564 in the low/hard state fit with (top
panel) one and (bottom panel) two thermal Comptonized continua and
their reflection.}
\label{fig:ls}
\end{figure}

The model is further constrained if the HEXTE data (30--200 keV) is
used as the electron temperature can be determined from the 
high energy rollover in the
spectrum. A single thermal Comptonization model (with its relativistic
reflection) can fit both PCA and HEXTE data ($\chi^2_\nu=71.5/75$), as
shown in the upper panel of figure \ref{fig:ls}. The reflection
parameters remain similar to those derived from the PCA data alone,
$R_{in}=26^{+48}_{-14}$, $\Omega/2\pi=0.25^{+0.04}_{-0.03}$ and
$\xi=14^{+26}_{-8}$. However, the broad
bandpass covered also shows evidence for subtle spectral
curvature which can be modelled by including a second Comptonized
component. This gives a significantly better fit and is
shown in the lower panel ($\chi^2_\nu=51.5/72$) but for rather
different reflected parameters ($\Omega/2\pi=0.0825^{+0.05}_{-0.05}$,
$\xi=352^{+9000}_{-280}$ and $R_{in}=20^{+31}_{-10}\, \Rg$). The second
Compton component has a shape which is clearly rather similar to a
reflection hump, so if this is the shape of the continuum 
then it significantly reduces the derived reflected fraction. 

While this spectral curvature is significantly detected it does not
necessarily mean that the continuum is truly described by a two (or
multi--) temperature form. Firstly, it could artifact of residual
cross--calibration uncertainties between the PCA and HEXTE instruments
(although the spectral curvature is marginally detected in the PCA
data alone: including a second Comptonized component (with its
reflection tied to that of the first Comptonized spectrum) gives a
reduction in $\chi^2_\nu$ to $27.3/34$). It could also indicate that
the reflected spectral models used here underestimate the Compton
reflection hump, such as might be expected if there was an ionized
skin overlying a neutral disk which produced Compton reflection but no
spectral features (Nayakshin, Kazanas \& Kallman 2000, Done \&
Nayakshin 2001). Alternatively there can be 
complex curvature in a Comptonized spectrum due to the visibility of
the individual scattering orders (e.g. Pozdnyakov, Sobol,\& Sunyaev 1983),
which are not included in our
approximate {\tt thComp} model. We use an exact solution of the Comptonization
equations for optically thin material ({\tt CompPS}: Poutanen \&
Svensson 1996). This does not include the self--consistent line
emission for the reflection spectrum, so the line is included as a
{\tt diskline} with free energy and normalisation. This gives a comparable 
fit to the PCA plus HEXTE data as the single {\tt thComp} model
$\chi^2_\nu=60.1/73$, with $\Omega/2\pi=0.26$ and $\xi=62$ i.e. for
similar reflected parameters to the single Compton component
model.

However the initial spectrum is modelled, it is very similar to
the classic Low State spectra seen in many Galactic Black Holes
(little or no Black Body and a large Compton component). The power
spectrum of these data shown by Cui et al. (1999) is also
very similar to low/hard
state power spectra (e.g. van der Klis 1995), being flat below $\sim 0.1$
Hz, breaking to $\sim f^{-1}$, and then to $\sim f^{-2}$ above $\sim 10$
Hz, with a high fractional {\it r.m.s.} amplitude of $\sim 30 \%$.
Comparing these with the compilation of RXTE Cyg X--1 data of Gilfanov,
Churazov \& Revnivtsev (1999) shows that both spectrum and variability are
typical of Cyg X--1 in its {\it hardest} low/hard state observed by RXTE. 

\begin{figure}
\epsfxsize=\hsize
\begin{center}
\epsfbox{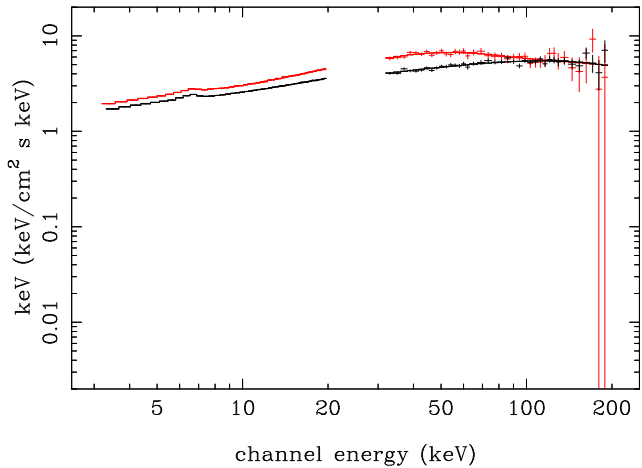}
\end{center}
\caption{The spectrum of J1550--564 in the low/hard state (above) overlaid 
with the hardest detected state of Cygnus X-1 (below).} 
\label{fig:cygx1}
\end{figure}

We illustrate this spectrally by showing the hardest low/hard spectrum of
the RXTE Cyg X--1 observation (10238-01-03-00), extracted identically to
the J1550--564 data. We fit this to the same spectral model of a disk
black body and two Comptonized continua (together with their reflected
emission). This gives an excellent fit, with $\chi^2_\nu=74.6/72$. Figure
\ref{fig:cygx1} shows this with the data from RXTE J1550--564 over-plotted.
Again, the spectrum is slightly curved: a single Compton continuum and its
reflection gives a significantly worse fit ($\chi^2_\nu=88.9/75$),
although again this could be due to the same cross-calibration
uncertainties rather than necessarily pointing to a more complex
continuum or reflected spectrum. 
Even the normalizations are similar, assuming that they are
both at similar distances (Cyg X--1 is at $\sim 2$ kpc: \Gierlinski et al.
1999). We conclude that in the first part of the rise then RXTE J1550--564
was in a classic LS. 

\subsection{Final spectrum}

The model which provided an acceptable fit to the LS PCA data (disk
blackbody as seed photons for thermal Comptonization and its reflected
spectrum with relativistic smearing) also gives an adequate fit to the
final spectrum ($\chi^2_\nu=44.4/37$) but with very different parameters.
The disk blackbody is now significantly detected, with $kT\sim 0.7$ keV,
and there is significant spectral curvature in the PCA spectrum which
causes the derived electron temperature to be $\sim 19$ keV, much
lower than the LS. The ionization state of the reflector is now
extremely high, $\xi\sim 3\times 10^3$. 
The phenomenological model (a disk black body and powerlaw) gives a
statistically similar fit,
$\chi^2_\nu=43.1/36$, with seed photon temperature $kT\sim 0.6$ keV.
Including the HEXTE data shows that the phenomenological model is
inadequate. A $\chi^2_\nu$ of $328.0/74$ is found although this drops to
$\chi^2_\nu=70.4/72$ with an exponential rollover in the continuum. 

\begin{figure}
\epsfxsize=\hsize
\begin{center}
\epsfbox{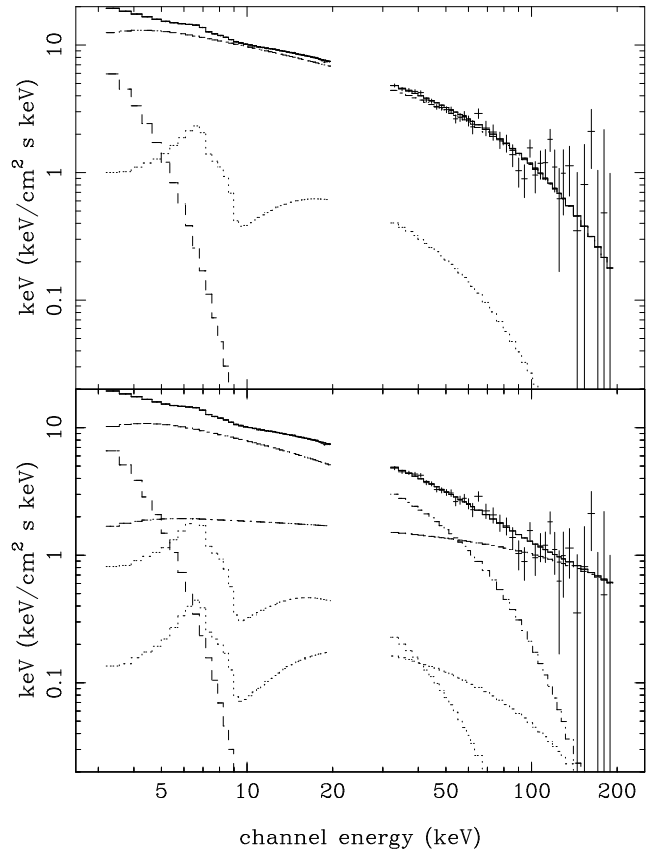}
\end{center}
\caption{The spectra of J1550--564 in the very high state fit with a disk 
black body and one thermal Compton component (top panel) and two
thermal Compton components (bottom panel).}
\label{fig:vhs}
\end{figure}

The single Comptonized component and its reflection (with a disk
blackbody from the accretion disk) can fit the PCA and HEXTE data
adequately ($\chi^2_\nu=89.1/75$), with $\Omega/2\pi = 0.11$,
$\xi=2.6\times 10^3$ and $R_{in}=15\, \Rg$. Once again there are hints
of spectral curvature.  Using a second Compton component again gives a
(marginally significantly) better fit ($\chi^2_\nu=79.43/72$) which is
shown in Figure \ref{fig:vhs}. However, this time the reflection
parameters are robust to changes in the spectral form, with
$\Omega/2\pi=0.12$, $\xi=2.5\times 10^3$ and $R_{in}=18\, \Rg$
for the two {\tt thComp} continuum, and 
$\Omega/2\pi=0.15$, $\xi=8.2\times 10^3$ and $R_{in}=13\, \Rg$ for a
single {\tt CompPS} continuum. 

The two temperature model continuum is gives a combination of low energy
curvature with a high energy tail which is very similar to that seen in Cyg
X--1 in its high state (Poutanen \& Coppi 1998; Coppi 1999;
\Gierlinski et al. 1999), the difference here being that low
temperature Comptonized emission is much stronger in RXTE J1550--564
(thus making the disk emission less obviously dominant) than in the HS
of Cyg X--1. This, together with its power spectrum (Cui et al. 1999)
clearly show that by the end of slow rise RXTE J1550--564 is in the
very high state. The classic high/soft state (defined as that which
contains a dominant black body spectrum and little or no Comptonized
component) is {\it never} encountered during the rise, and the
spectral transition is from LS to VHS.

\subsection{Intermediate Data}

\begin{figure}
\epsfxsize=\hsize
\begin{center}
\epsfbox{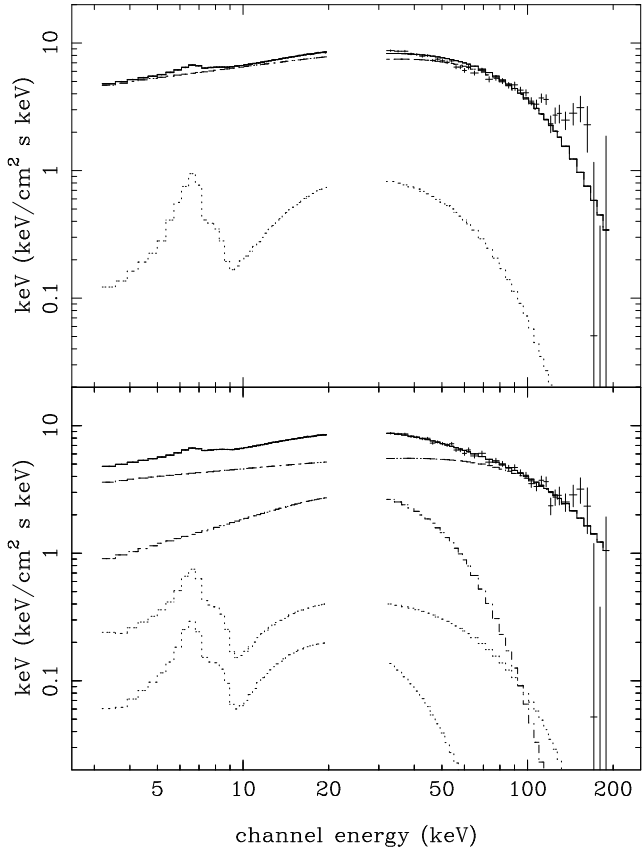}
\end{center}
\caption{The spectra of J1550--564 in an intermediate state fit with a disk 
black body and one thermal Compton component (top panel) and with two
thermal Compton components (bottom panel).}
\label{fig:is}
\end{figure}

Any system of models must be able to fit all the data, so here we look
at a spectrum taken from the middle of the rise (spectrum number 3),
where the hardness ratio clearly indicates that the source is in
transition.  A single Compton component model using the PCA data alone
gives $\chi^2_\nu=27.0/37$ with $\Omega/2\pi=0.07$, $\xi=7\times
10^3$, and $R_{in}=45\, \Rg$,
but including the HEXTE data gives $225.2/75$, a completely inadequate
fit. The curvature is {\it not} well modeled even by {\tt CompPS}
which gives $\chi^2_\nu=129/73$. The two component Comptonization model is
the only one which gives a good fit to the broad band data
($\chi^2_\nu=66/75$), as shown in Figure \ref{fig:is}. The reflection
parameters are then similar to those derived from the PCA data alone
with $\Omega/2\pi=0.06$, $\xi=7.1\times 10^3$, and $R_{in}=41\, \Rg$. The 
nature of the spectral
curvature is now not at all subtle, and the spectrum clearly shows
that there is a low temperature, curving component in the PCA,
together with rather higher temperature emission which dominates the
HEXTE band. In the final spectrum the lower temperature component is
somewhat hotter and dominates more of the high energy spectrum, making
the single temperature models an adequate description.

\subsection{All Data}

We are interested in how the reflected spectrum varies during the
transition, yet the previous sections have shown that these are not
always robust to changing the continuum spectral form. Accordingly we
give results for two set of spectral fits. In the first we assume that
the HEXTE cross--calibration is reliable and fit all the rise
PCA/HEXTE spectra with a model consisting of seed disk blackbody
photons, together with two thermal Compton continua and their
relativistically smeared reflection (Table 1). The last column of
Table 1 gives the $\chi^2_\nu$ for a single Compton continuum model to
the PCA/HEXTE data, showing that it cannot acceptably fit most of the
intermediate spectra (see previous section).  The second set of fits
assumes that the HEXTE cross--calibration is not reliable at this
level and so uses only the PCA data with a model of seed disk
blackbody photons, a single temperature thermal Compton continuum
and its relativistically smeared reflection (Table 2).  Given the
uncertainty on the continuum model (and the fact that the temperatures
derived from {\tt thComp} are inaccurate above $\sim 50$ keV) we show
spectral evolution by the changing hardness ratio of the PCA data.

For the single temperature continuum fits to the PCA data the onset of
the LS--VHS transition is marked by an increase in ionization state of
the reflector from $\sim 20$ to $\sim 5000$ and a decrease in the
solid angle subtended by the disk, from $\Omega/2\pi \sim 0.25-0.12$.
Apart from this jump at spectrum 3 the parameters are remarkably
stable throughout large changes in spectral form. This stability is
even more marked in the two temperature PCA/HEXTE fits. There is some
indication of a jump in ionization at spectrum 3, at the onset of the
LS--VHS transition, though it is less significant due to the larger
error bars on the reflection parameters from the LS data. However,
there is no evidence for any change in the solid angle.

\begin{figure}
\epsfxsize=\hsize
\begin{center}
\epsfbox{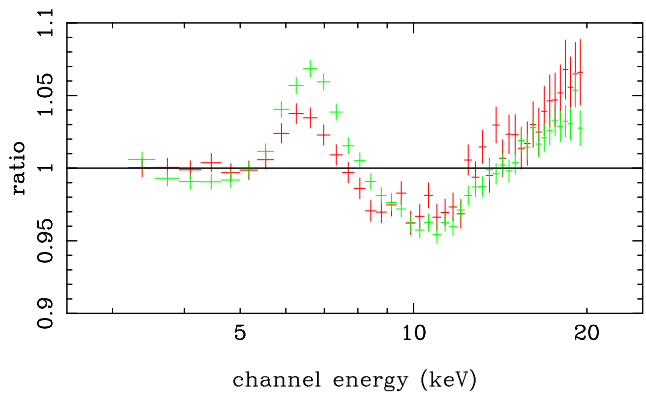}
\end{center}
\caption{The residuals to a continuum only model fit to data before
(first spectrum) and after (fifth spectrum) the shift in reflection
parameters. Plainly the line and edge energy increase, but only the
line increases in strength, while the edge depth and Compton hump stay
approximately constant.} \label{fig:ratio}
\end{figure}

Ionization and the amount of reflection are 
correlated in the fits.  As the ionization increases the photoelectric
opacity decreases, so there is more reflected continuum below $\sim
15$ keV. The spectral features (line and edge) shift to higher
energies, and their contrast with respect to the continuum increases
(although this also depends on the hardness of the illuminating
continuum). In general the line increases by more than the edge as the
fluorescence yield increases with ionization state (see e.g. Ross \&
Fabian 1993; \Zycki \ \& Czerny 1994 for photo--ionized reflected
spectra). Figure \ref{fig:ratio} shows the residuals to the best fit
continuum only model (disk blackbody and single thermal Compton
spectrum but no reflection) to the data before and after the
transition (PCA spectra 1 and 5). There is clearly more line and its
mean energy has shifted up, indicating ionization. The edge energy has
also increased, but its strength has not. While ionization predicts
that the line should increase by more than the edge, the edge depth
should also increase if the solid angle remains the same. So to get
increasing ionization with the same edge depth requires that the solid
angle decreases.

\begin{table*}
\begin{minipage}{180mm}
\caption{The table shows the reflection parameters for the PCA and HEXTE 
data using a model of a disk black body with two Comptonized 
reflectors. The $\chi^2_\nu$ values for a single Compton reflector are
also included. Errors are calculated for $\Delta\chi^2=2.7$} 
\label{}
\begin{tabular}{lcccccc}

\hline

Spectrum number & $\Omega/2\pi$\footnote{Solid angle subtended by 
reflector} & $\xi$\footnote{Ionization parameter} & Inner 
radius \footnote{Measured in gravitational radii, $GM/c^2$} & hardness 
ratio \footnote{Ratio of 8-20keV and 1-5keV data} & $\chi^2_\nu$ \footnote{For
two Compton components} & $\chi^2_\nu$ \footnote{For one Compton
component} \\

\hline

1 & $0.08^{+0.05}_{-0.05}$ & $350^{+9150}_{-280}$ & 
$20^{+31}_{-10}$ & $2.05^{+0.05}_{-0.05}$ & $51/72$ & $72/75$ \\

2 & $0.10^{+0.04}_{-0.025}$ & $375^{+275}_{-275}$ & 
$21.5^{+23.5}_{-8.5}$ & $1.95^{+0.05}_{-0.05}$ & $65/72$ & $104/75$ \\

3 & $0.07^{+0.01}_{-0.006}$ & $5100^{+4700}_{-2400}$ & 
$42^{+73}_{-19}$ & $1.85^{+0.05}_{-0.05}$ & $66/72$ & $225/75$ \\

4 & $0.08^{+0.01}_{-0.007}$ & $4800^{+4220}_{-3280}$ & 
$30.5^{+52}_{-13}$ & $1.82^{+0.03}_{-0.03}$ & $59/72$ & $147/75$ \\

5 & $0.09^{+0.01}_{-0.006}$ & $4480^{+3920}_{-2280}$ & 
$28^{+23}_{-12}$ & $1.60^{+0.05}_{-0.05}$ & $59/72$ & $192/75$ \\

6 & $0.08^{+0.01}_{-0.009}$ & $3865^{+3135}_{-1965}$ & 
$49^{+226}_{-23}$ & $1.50^{+0.03}_{-0.03}$ & $62/72$ & $165/75$ \\

7 & $0.10^{+0.02}_{-0.008}$ & $2230^{+1470}_{-1430}$ & 
$24^{+18}_{-9.5}$ & $1.25^{+0.025}_{-0.025}$ & $63/72$ & $166/75$ \\

8 & $0.11^{+0.01}_{-0.01}$ & $2190^{+1800}_{-1100}$ & 
$26^{+23}_{-10}$ & $1.00^{+0.03}_{-0.03}$ & $91/72$ & $168/75$ \\

9 & $0.12^{+0.02}_{-0.01}$ & $2375^{+1525}_{-1480}$ & 
$19^{+11}_{-6}$ & $0.85^{+0.01}_{-0.01}$ & $69/72$ & $118/75$ \\

10 & $0.12^{+0.02}_{-0.01}$ & $2450^{+2550}_{-1450}$ & 
$17.5^{+12.5}_{-7}$ & $0.87^{+0.015}_{-0.015}$ & $58/72$ & $90/75$ \\

11 & $0.12^{+0.015}_{-0.01}$ & $2520^{+1680}_{-2040}$ & 
$18^{+15}_{-5.5}$ & $0.87^{+0.01}_{-0.01}$ & $81/72$ & $117/75$\\

12 & $0.12^{+0.03}_{-0.01}$ & $2200^{+1400}_{-1600}$ & 
$17^{+11}_{-7.5}$ & $0.77^{+0.015}_{-0.015}$ & $72/72$ & $92/75$ \\

13 & $0.12^{+0.02}_{-0.01}$ & $2400^{+3575}_{-1400}$ & 
$19.5^{+17.5}_{-7}$ & $0.90^{+0.01}_{-0.01}$ & $68/72$ & $94/75$ \\

14 & $0.12^{+0.015}_{-0.01}$ & $2500^{+3500}_{-1390}$ & 
$18^{+13}_{-7.5}$ & $0.75^{+0.01}_{-0.01}$ & $80/72$ & $89/75$ \\

\hline
\end{tabular}
\end{minipage}
\end{table*}

\begin{figure}
\epsfxsize=\hsize
\begin{center}
\epsfbox{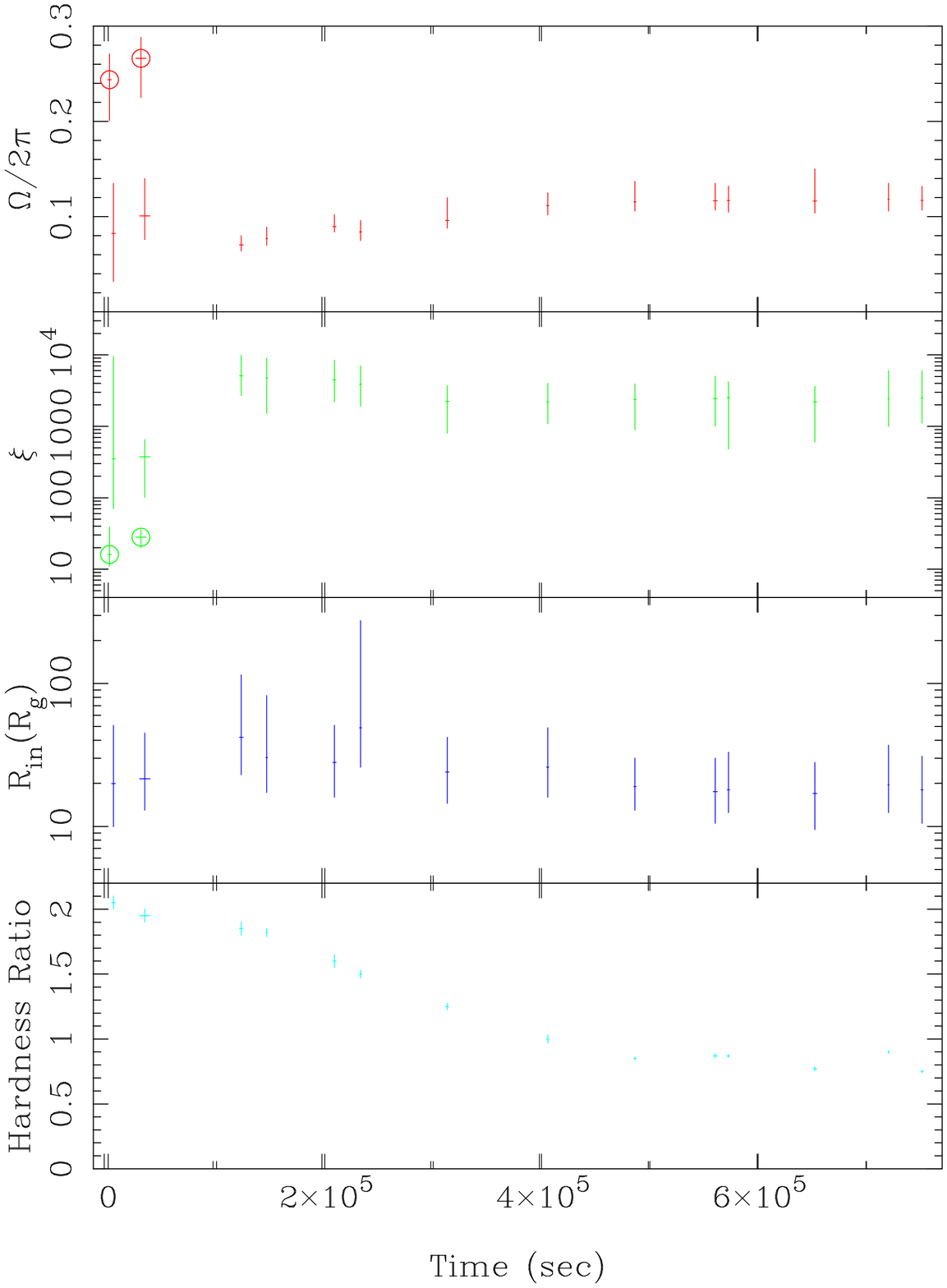}
\end{center}
\caption{The spectral transition of J1550--564 represented as a change
in hardness ratio.  Also included are the changes in covering
fraction, ionization state and disk inner radius.}
\label{fig:reflect}
\end{figure}

The evolution of the reflector parameters during the rise is shown in
Figure \ref{fig:reflect}. Irrespective of the continuum model used,
the derived amount of reflection is always significantly less than
unity, while the amount of relativistic smearing (while always
significant although with large error bars) is {\it never} as large as
that expected for a disk extending down to the last stable orbit in a
Schwarzchild geometry.

If the single Comptonized component is the correct model for the
hard spectrum in the low state then there is a sharp increase in
ionization state, coupled with a decrease in the amount
of reflection at the onset of the spectral state transition. 
In the two component Comptonization model then there are 
hints of the same effect, although it is less significant due to the
larger error bars on the derived ionization state. 

\begin{table*}
\begin{minipage}{180mm}
\caption{The table shows the reflection parameters for the PCA data
only, using a model of a disk black body with a single Comptonized 
reflector. Errors are calculated for $\Delta\chi^2=2.7$} 
\label{}
\begin{tabular}{lccccc}

\hline

Spectrum number & $\Omega/2\pi$\footnote{Solid angle subtended by 
reflector} & $\xi$\footnote{Ionization parameter} & Inner 
radius \footnote{Measured in gravitational radii, $GM/c^2$} & hardness 
ratio \footnote{Ratio of 8-20keV and 1-5keV data} & $\chi^2_\nu$
\footnote{For one Compton component} \\

\hline

1 & $0.24^{+0.03}_{-0.04}$ & $16^{+23}_{-5}$ &
$26.5^{+63.5}_{-15}$ & $2.05^{+0.05}_{-0.05}$ & $37/37$ \\

2 & $0.27^{+0.02}_{-0.04}$ & $28^{+7}_{-8}$ &
$28.5^{+33}_{-11}$ & $1.95^{+0.05}_{-0.05}$ & $44/37$ \\

3 & $0.07^{+0.01}_{-0.01}$ & $7250^{+4700}_{-3750}$ &
$45^{115+}_{-20}$ & $1.85^{+0.05}_{-0.05}$ & $27/37$ \\

4 & $0.08^{+0.02}_{-0.01}$ & $5345^{+5000}_{-3200}$ &
$33^{+53}_{-16}$ & $1.82^{+0.03}_{-0.03}$ & $22/37$ \\

5 & $0.09^{+0.01}_{-0.01}$ & $5300^{+5000}_{-2700}$ &
$28^{+25}_{-10.5}$ & $1.60^{+0.05}_{-0.05}$ & $34/37$ \\

6 & $0.09^{+0.01}_{-0.01}$ & $4750^{+3330}_{-2360}$ &
$50.5^{+180}_{-24.5}$ & $1.50^{+0.03}_{-0.03}$ & $30/37$ \\

7 & $0.09^{+0.03}_{-0.01}$ & $2100^{+2000}_{-1380}$ &
$24^{+18}_{-10.5}$ & $1.25^{+0.025}_{-0.025}$ & $25/37$ \\

8 & $0.11^{+0.01}_{-0.02}$ & $2380^{+2320}_{-1280}$ &
$23.5^{+23}_{-8.5}$ & $1.00^{+0.03}_{-0.03}$ & $41/37$ \\

9 & $0.11^{+0.02}_{-0.03}$ & $2290^{+2200}_{-1540}$ &
$27^{+24}_{-16}$ & $0.85^{+0.01}_{-0.01}$ & $41/37$ \\

10 & $0.12^{+0.02}_{-0.01}$ & $2280^{+3000}_{-1200}$ &
$23^{+15.5}_{-12}$ & $0.87^{+0.015}_{-0.015}$ & $37/37$ \\

11 & $0.11^{+0.02}_{-0.01}$ & $2325^{+3675}_{-1475}$ &
$26^{+22}_{-15}$ & $0.87^{+0.01}_{-0.01}$ & $35/37$ \\

12 & $0.12^{+0.02}_{-0.02}$ & $1850^{+1650}_{-1000}$ &
$19^{+20}_{-8.5}$ & $0.77^{+0.015}_{-0.015}$ & $36/37$ \\

13 & $0.12^{+0.02}_{-0.01}$ & $2300^{+5200}_{-1500}$ &
$25^{+23}_{-13}$ & $0.90^{+0.01}_{-0.01}$ & $34/37$ \\

14 & $0.12^{+0.02}_{-0.02}$ & $2500^{+3500}_{-1300}$ &
$19^{+20}_{-8}$ & $0.75^{+0.01}_{-0.01}$ & $44/37$ \\

\hline
\end{tabular}
\end{minipage}
\end{table*}

\section{Models of the low state during the outburst}

As has been mentioned, there are two main theories for how the LS spectrum
is formed. The first is that the accretion disk truncates, forming an
inner, optically thin, X--ray hot accretion flow, while the second
postulates magnetic flares above a disk. We review here the very
different predictions of these models for how the X--ray emitting
region evolves during the rise to outburst, but show that both can 
give the observed LS spectrum. 

\subsection{Truncated disk with an inner, X--ray hot flow}

The geometry of a truncated disk and inner, X--ray hot flow 
can plainly explain the weak disk emission, hard (photon--starved)
X--ray spectrum, and generally low amplitude of X--ray reflection seen
in the LS (e.g. Poutanen, Krolik \& Ryde 1997). 
Certainly in quiescence it seems very likely that the disk
structure does not extend all the way down to the last stable
orbit. Firstly, the observed X--ray luminosity, while low, still
implies a mass accretion rate which is too high for the inner disk to
remain below the instability threshold i.e. to be quiescent. A disk
which truncates at larger radii can carry this mass accretion rate
without triggering the instability (Lasota et al. 1996). Also, the
inner regions of a quiescent disk structure are generally
optically thin to both true absorption and electron scattering (see
e.g. Cannizzo 1998). Such material cannot cool efficiently, so will
heat up. Conduction between this and the remaining quiescent disk can
lead to progressive evaporation of the inner regions of the quiescent
disk into a hot flow (Meyer \& Meyer--Hofmeister 1994;
Meyer--Hofmeister \& Meyer 1999). Conduction from a hot flow can lead
to evaporation of even an SS disk, so this can be a mechanism to
truncate the disk even at the much higher mass accretion rates
generally seen for the LS emission (\Rozanska \& Czerny 2000).
The only known potentially stable solution of the accretion equations
for an optically thin hot flow are the Advection Dominated Accretion
Flows (ADAFs: Narayan \& Yi 1995). 

When the instability is triggered in the outer quiescent disk then
this brings a large amount of mass in towards the black hole which
increases the ADAF's density, so increasing its luminosity.  However,
at a certain critical mass accretion rate, the density is such that
electron--ion collisions become frequent, so the electrons gain most
of the ion energy. The advected fraction becomes small and the ADAF
collapses into a standard SS disk. This predicts that the LS will end
at accretion rates of $\dm\sim 1.3\alpha^2$ (Where $\dm = {\dot
M}/{\dot M_{Edd}}$, and ${\dot M_{Edd}}=10 L_{Edd}/c^2 $). This  corresponds to luminosities of $1-8 \%$ of Eddington for a
viscosity of $\alpha =
0.1-0.3$, as around a third of the energy is advected rather than
radiated at these densities (Esin et al. 1997; Quataert \& Narayan
1999).  At these high viscosities the flows are also stable to
convection (Narayan, Igumenshchev \& Abramowicz 2000).

These models predict $\Omega/2\pi < 1$ as the ADAF is centrally
concentrated, so the solid angle subtended by the disk is fairly small
(Esin et al. 1997). However, more detailed predictions are difficult
as they critically depend on the geometry at the transition between
the disk and ADAF. Current models generally {\it assume} a transition
radius of $\sim 1000$ Schwarzchild radii, and these give LS spectra
which have $\Omega/2\pi << 1$ (Esin et al. 1997) and very little
relativistic smearing. Previous observations of 2--20 keV LS spectra 
show  $\Omega/2\pi \sim 0.3$ (\Gierlinski et al. 1997; \Zycki \
et al. 1997; 1998; Gilfanov et al. 1999), and smearing characteristic
of an inner radius of 10--20 Schwarzchild radii (\Zycki \ et al. 1997;
1998; 1999; Done \& \Zycki \ 1999). 
These are incompatible with a disk truncated at 1000 Schwarzchild radii.
While both the solid angle and smearing can match the observations 
by simply decreasing the transition radius between the ADAF and the
cold disk, this is not a completely free parameter. A small transition
radius gives rise to a strong disk flux which Compton cools the ADAF,
and can lead to its collapse (Esin 1997). A transition at 10
Schwarzchild radii decreases the maximum ADAF luminosity by a factor
of $\sim 2$, making it only just feasible to get the observed LS
luminosity of $\sim 3 \%$ of Eddington (Esin et al. 1997; \Gierlinski et
al. 1999). However, these problems disappear entirely 
if the PCA/HEXTE cross--calibration spectra is reliable. Continuum curvature 
can give an {\it overestimate} of the amount of reflection present in 
single temperature/power law fits to low energy ($\le 20$ keV)
spectra (compare Tables 1 and 2 for the LS spectra).

The inner radius of most of the disk material is almost certainly
moving during the rise to outburst, If the edge of the ADAF follows
the inner radius of the heating wave (i.e. the inner radius of the SS
part of the disk), then the reflected fraction should be small and
remain fairly constant. Alternatively, if the disk penetrates some way
into the ADAF, $\Omega/2\pi$ is rather larger and can increase or
remain constant as the heating wave moves inwards, depending on the
details of the geometry. Both of these options are consistent with our
derived model parameters, as our data do not cover much of the LS
evolution.

While this gives continuity of properties between the quiescent and LS
spectra, there should be an abrupt change when the ADAF finally collapses
into whatever very different mechanism powers the LS/VHS
emission. The
viscous time-scale for a quasi--spherical flow to collapse is only $\sim
1/\alpha\times$ longer than the dynamical time-scale at a given radius i.e.
much less than a second for a flow of $10 R_s$ around a $10\MSun$ black
hole. However, the collapse might be triggered locally by the disk
underneath the hot flow, in which case it will be determined by the
viscous time-scale for the SS disk, which is $\sim$ a few hours for the
same parameters.  This is still a much shorter than the observed $\sim 3$
day time-scale for the spectral transition (see Figure 1), although
evaporation of the disk material into the hot flow may slow this
considerably (\Rozanska \& Czerny 2000). 

\subsection{Magnetic corona above an untruncated disk}

In the second theory for the LS emission, some (large) fraction of the
gravitational potential energy of the infalling material is released
in an optically thin environment by magnetic reconnection above the
dense disk material. In quiescence this can power the observed low
level X--ray emission without triggering the instability as long as
these are distributed over the whole disk (powered by the local
$\mdot$ at a given radius which is {\it not} constant) rather than
requiring that the X--rays are produced solely by the central $\mdot$
(Nayakshin \& Svensson 2001). However, numerical simulations imply that the
Balbus--Hawley dynamo mechanism for the magnetic field shuts off when
the disk material becomes mainly neutral (so has large resistivity:
Gammie \& Menou 1998; Fleming, Stone \& Hawley 2000). The quiescent
disk is then unlikely to be able to power {\em any} magnetic
reconnection, so the continuity of spectral properties from the LS to
quiescence is hard to explain.  Also, the inner quiescent disk is
predicted to be optically thin to both absorption and scattering, so
the caveats listed in the previous section still apply as to whether
it can exist as a geometrically thin, cool disk.

Irrespective of whether this can produce the hard X--rays observed in
quiescence, the X--ray emission from magnetic reconnection above the
outbursting disk is likely to be dominant once the the instability is
triggered since its local $\mdot$ will be so much larger. The heating wave
starts in the outer disk and moves inwards, carrying with it an ever
increasing amount of matter. While the disk would remain quiescent in
front of the wave, it would switch into a hotter SS state behind it (e.g.
Cannizzo 1998).  The X--rays during the rise are then associated with
magnetic reconnection generated by the Balbus--Hawley mechanism above the
outbursting disk.  In this case we would expect to see a steadily
decreasing radius (i.e. an increasing amount of relativistic smearing) and
$\Omega/2\pi$ remaining constant at a value of unity. 

LS spectra in general (both in Cyg X--1 and during the
decline phase of the transient systems) show $\Omega/2\pi<
1$ (irrespective of the continuum model used), 
which is incompatible with the magnetic flares model described
above (\Gierlinski et al. 1997; \Zycki, Done \& Smith 1997; 1998; Done
\& \Zycki \ 1999; Gilfanov et al. 1999; Zdziarski, \Lubinski \& Smith
1999). However, if the reconnection regions are out-flowing rather than
static then the reflection signature can be suppressed by the resulting
anisotropic emission (Beloborodov 1999). For a constant outflow
velocity then this predicts that $\Omega/2\pi$ remains constant at
some value $< 1$, while the amount of relativistic smearing
increases. Alternatively, the generally observed LS correlation
between spectral index and $\Omega/2\pi$ can be explained by a
variable outflow velocity.  Faster outflow velocities mean that more
of the X--ray radiation is beamed away from the disk so there is a
smaller reflected fraction. This also means fewer soft photons from
the disk are intercepted by the active region, so the spectra are
harder (Beloborodov 1999). The observed softening as the outburst
progresses then requires that the outflow speed decreases
systematically through the rise (as shown in figure
\ref{fig:corona2}). Interestingly, this gives the observed correlation
between jet strength and spectral state if the outflow is treated as
the base of a radio jet (Fender 2000).

\section{Models of the very high state during the outburst}

The (admittedly few) broad band HS/VHS spectra show a rather complex
spectrum. There is plainly a strong disk component at $\sim 1$ keV,
and a high energy tail.  The VHS spectrum shown in Figure
\ref{fig:vhs} shows the tail extending out beyond 200 keV,
while OSSE data from other objects show it out to even higher energies
(Grove et al. 1998; \Gierlinski et al. 1999).  The lack of distinct
scattering orders strongly argues against thermal Comptonization
models in the HS of 
Cyg X--1 (\Gierlinski et al. 1999), and the two proposed mechanisms in
the literature involve a non--thermal electron distribution either
from magnetic flares (Poutanen \& Coppi 1998) or from bulk motion of
the infalling material as it approaches the black hole event horizon
(Chakrabarti \& Titarchuk 1995).  The major problem with the bulk
motion model is that the free--fall electron velocities are not high
enough (typical velocities of $\sim 0.7c$ imply Lorentz factors of
only $\gamma\sim 1.4$) to extend the power law past $\sim 100-300$ keV
(Laurent \& Titarchuk 1999, Zdziarski 2000) yet the highest
signal--to--noise HS/VHS spectra extend unbroken beyond this (GRO
J1655-40: Grove et al. 1998; Cyg X-1: \Gierlinski\ et al., 1999). Thus
it seems that the power law {\it must} arise from highly relativistic
non--thermal electron distribution. 

As well as the non--thermal power law, there is an additional
continuum component which is clearly seen in the few broad band VHS/HS
spectra (Figure \ref{fig:vhs}, Figure \ref{fig:is} 
and Cyg X--1: \Gierlinski\ et al. 1999;
Frontera et al. 2000). This can be described as a thermal Comptonized
continuum, but at lower temperature than the standard LS emission. It
seems entirely possible that the difference between the VHS and HS
spectra is merely in the strength of this additional thermal
component. This thermal/non--thermal (or hybrid) plasma most likely 
represents incomplete thermalization of the magnetic reconnection
energy, so that the thermal and non--thermal electron distributions
are co--spatial rather than there being two separate electron
populations (Poutanen \& Coppi 1998; Coppi 1999).

\begin{figure*}
\epsfxsize=\hsize
\begin{center}
\epsfbox{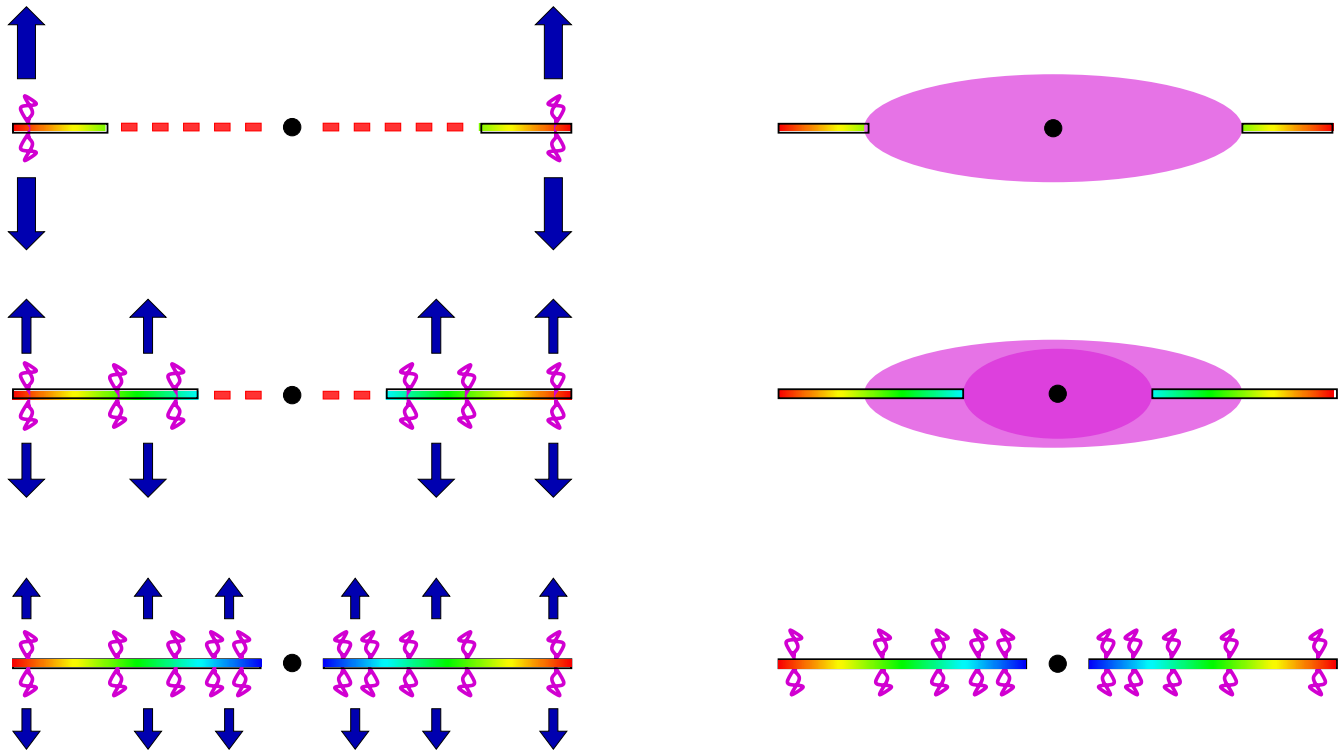}
\end{center}
\caption{A representation of the quiescent disk and ADAF models. The left
hand side shows a quiescent disk being gradually replaced by a hot/thin
disk as the outburst progresses. A slowing relativistic wind is theorized
as a mechanism for reducing the reflected fraction as the outburst
progresses. The Balbus-Hawley reconnection regions are only present in the
hot disk. The right hand side shows an ADAF being penetrated and
eventually destroyed by an encroaching hot disk. In the final state,
magnetic reconnection acts as the hard X--ray source.}
\label{fig:corona2}
\end{figure*}

Since there is very little radio emission  in the HS (Fender 2000)
then the emission regions are probably not linked to a relativistic 
outflow. Thus the HS (and by extension the VHS) are most likely to be linked
with rather low velocity outflows, or static regions above the disk, 
as shown schematically in Figure \ref{fig:corona2}, so should give
rise to $\Omega/2\pi\sim 1$ in the reflector, and to an inner radius
equal to the last stable orbit.

\section{Discussion}

The parameters which we actually derive are shown in figure
\ref{fig:reflect}. The reflected fraction, $\Omega/2\pi$, either
remains essentially constant around a value of $\sim 0.1$, or drops
from $\sim 0.25$ to $0.1$ at the onset of the LS--VHS transition
(depending on the continuum model used). The inner radius remains
fairly constant: the errors on the data are large enough to cover most
trends although we note that the results are never consistent with a
disk extending down to the last stable orbit, nor with a disk
truncated at 1000 Schwarzchild radii.

The LS spectrum is {\it roughly} consistent with the presence of
either an ADAF or of out-flowing magnetic flares above the
(outbursting) disk. The data can even be compatible with {\it static}
magnetic flares, as there is an ionization instability which is
present in X--ray illuminated material in hydrostatic balance. Intense
X--ray irradiation of the disk can produce a sharp transition in the
vertical structure of the disk, with a highly ionized skin forming on
top of mainly neutral material. In the LS, the hard spectrum has a
Compton temperature which is high enough to completely strip iron.
The skin is then almost completely reflective, and forms no spectral
features. The observed reflected signature is then dominated by
photons reflected from deeper in the disk, where the material is much
less ionised (Nayakshin, Kazanas \& Kallman 2000). If the skin has an
optical depth of around unity then this appreciably reduces the
reflection signature to $\Omega/2\pi < 0.3$ (Nayakshin 2000, Done \&
Nayakshin 2001).

For the ADAF, the presence of a skin on the top of the disk could give
rise to an {\it overestimation} of the amount of relativistic
smearing. The observed, mainly neutral, reflected spectrum is subject
to Compton scattering in the skin before it escapes. This can lead to
some broadening of the spectral features (line and edge) in addition
to the relativistic broadening of the features (see also Ross, Fabian
\& Young 1999). Thus the theoretical
difficulties in maintaining an ADAF with a small transition radius
could be removed if the transition radius is overestimated because of
Compton smearing.  With magnetic flares, the ionized skin gives the
observed $\Omega/2\pi <1$ and correlation with spectral index, without
requiring a moderately relativistic outflow from the disk (although
such outflow may be additionally present!).

The detailed properties of the VHS reflected spectra are not at all
consistent with the magnetic flares model developed earlier. The
amount of reflection is very much less than $\sim $ unity predicted by
such flares at low outflow velocity. However, the ionization is also
high, indicating that we are {\it not} dealing with a mainly neutral
disk.  Unlike the LS spectrum, photo--ionization of the disk surface
cannot easily suppress the amount of observed reflection. The Compton
temperature is low, so the ionization state of the X--ray illuminated
skin is not high enough to completely strip iron, and the derived
solid angle is generally not strongly underestimated (Nayakshin,
Kazanas \& Kallman 2000; Done \& Nayakshin 2001). However, there can
also be substantial {\it collisional} ionization as the observed disk
temperature is $\sim 0.7$ keV, provided that the disk is in local thermal
equilibrium (a likely assumption). We used reflection models which
assume that the whole disk can be described by a single {\it
photo--ionization} parameter, but plainly the hotter the disk, the
more highly ionized it will be simply due to collisional
processes. From the Saha equation the mean ionization state of iron
will be one in which the ionization energy is roughly equivalent to
the temperature i.e. we expect that iron should be dominated by He--
and H--like ions in the central regions where the X--rays are
predominantly produced. However, the electrons are very nondegenerate,
and a better estimate of the ionization state is one with ionization
energy $\sim 20 kT$ i.e. the inner regions of the disk will contain
iron  that has been completely stripped through collisional ionization
processes (Rybicki \& Lightman 1979). Thus reflection from the inner disk
is completely ionized, and has no spectral features. It is, therefore,
counted as continuum rather than reflected flux in spectral
fitting. Observable reflection is only detected from further out in
the disk, where relativistic smearing effects are smaller, although
Compton up-scattering can again be an important broadening mechanism
(Ross, Fabian \& Young 1999) which is not accounted for in our
reflection code.

\begin{figure}
\epsfxsize=\hsize
\begin{center}
\epsfbox{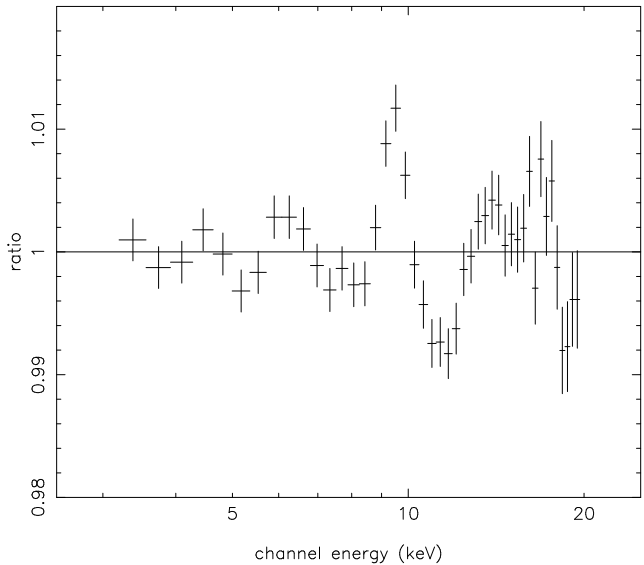}
\end{center}
\caption{Co-added residuals of the later rise of J1550--564.}
\label{fig:res}
\end{figure}

The importance of Compton up-scattering on our spectra can be
demonstrated by the residuals to the VHS spectra.  The co--added
residuals to all the spectra after the sudden parameter transition are
shown in figure \ref{fig:res}.  There is clearly a systematic problem
with the modelling of the reflected features at a level of $\sim
1\%$. A similar pattern of residuals is seen when models of ionized
disks which include Compton up-scattering of the reflected spectrum are
fit with the simpler reflection models used here (S. Nayakshin,
private communication).

Complete collisional ionization of the inner disk in the VHS can then
lead to an underestimate of the solid angle subtended by the
reflector, and to an overestimate of the inner disk radius, although
this latter effect can be partially compensated for by the Compton
up-scattering associated with the ionized disk reflection.  Thus our
derived $\Omega\sim 0.1$ and inner radius $\ge 10$ Schwarzchild radii
may be consistent with a disk which subtends a solid angle of $\sim 1$
and extends down to the last stable orbit when {\it collisional}
ionization effects are properly included.

This leaves us in the unfortunate situation of knowing even less than
we did before. Ionization removes the last of our ability to test
models for the origin of the X--ray spectra using currently
available spectral models for fitting the X--ray reflected spectrum.  
However, we note one
inconsistency which we suspect will remain even within properly
calculated ionization models.  The continuing increase in frequency of
the QPO indicates that the switch to the VHS occurs before the heating
wave has propagated all the way down to the last stable orbit. If so
then we expect that the magnetic corona will continue to increase in
luminosity as the disk radius decreases, but without a corresponding
reflected signature. This would imply that the hard X--ray flux should
increase, without a corresponding increase in reflection, so the
measured reflected fraction should decrease (while the ionization
state and relativistic smearing remain constant). However, we see the
solid angle remain constant from the onset of the VHS
transition. Plainly there is much that is still not understood about
the nature of the VHS emission. 

\section{Conclusions}

We have intensively studied the spectra of the rise phase of the
outburst of RXTE J1550--564. At the onset of the XTE coverage, the
source is in a classic LS, and can be fit with a spectrum consisting
of seed photons from an accretion disk (not detected in the RXTE
bandpass) being thermally Comptonized into a hard X--ray spectrum
which is then reflected from the disk.  There is subtle curvature in
the spectrum, perhaps indicating that a single temperature
Comptonization model is inadequate, although it may also be an
artifact of remaining residuals in the PCA/HEXTE
cross--calibration. The source then makes a {\it very} smooth transition to
the VHS (without going through the classic HS), where the spectrum is
dominated by disk emission which is strongly Comptonized. The high
energy spectral curvature during the transition is inconsistent with a
single temperature Comptonization model. This effect is much larger in
the later spectra and so is unlikely to be due to calibration
issues. The transition and VHS spectra can be modeled by two thermal
Compton components (one at low electron temperature, the other
considerably higher). This could be indicative of a hybrid
thermal--nonthermal plasma such as has been fit to Cyg X--1 HS spectra
(\Gierlinski\ et al. 1999).

We are most interested in the reflected spectrum rather than the
continuum, as this can be used to give an indication of the source
geometry and whether this changes dramatically at the LS/VHS
transition as required by ADAF models. 
We have derived values for the reflected fraction, the ionization
parameter and the inner disk radius for each spectrum.

A sharp transition is seen in the ionization
at the onset of the LS--VHS transition. The reflected fraction remains
essentially constant at a value of 
$\Omega/2\pi =0.1$ while the ionization increases from mainly
neutral to highly ionized (iron in He and H--like ionization states).
The inner radius inferred from relativistic smearing of the reflected
spectral features could not be well constrained, but is never as small
as the last stable orbit around a black hole. 

Given theoretical uncertainties in the transition radius and extent of
overlap between the ADAF and disk, then the LS reflected spectrum is
consistent with an ADAF interpretation, especially if
photo--ionization of the disk gives rise to an ionized skin which
introduces significant Compton broadening as well as relativistic
smearing. Similarly, a photo--ionized skin can make magnetic
reconnection models of the X--ray flux fit the data, even without
significant outflow velocities of the hard X--ray region, although
these can be additionally present.  However, the fact that the LS is
seen when the increasing QPO frequency and disk instability models
clearly indicate that the outbursting disk has not yet reached the
last stable orbit shows that an inner SS disk is not a {\it necessary}
condition on the LS X--ray emission.

Ionization can again distort the reflected spectrum derived in the
VHS, only this time it is collisional rather than photo ionization
which is likely to be important (Given a disk which is in LTE). Again
this can suppress the observed
amount of reflection, and distort (generally suppress) the most highly
relativistically broadened features from the inner disk. 

\section{Acknowledgements}

We thank Piotr \Zycki\ for the use of his Comptonization and
reflection codes and Sergei Nayakshin for illuminating
discussions. We also thank our referee, Andrzej Zdziarski, for his
helpful comments. CDW acknowledges support from a PPARC studentship.
This research has made use of data obtained through the High Energy
Astrophysics Science Archive, provided by the NASA Goddard Space
Flight Center.

{}

\end{document}